\documentclass{article}
\usepackage{graphicx} 
\usepackage[usenames,dvipsnames,svgnames,table]{xcolor}

\RequirePackage[T1]{fontenc} \RequirePackage[tt=false, type1=true]{libertine} 

\usepackage[a4paper, total={6.5in, 9.5in}]{geometry}
\usepackage{amsmath,amsthm,amssymb}
\usepackage[dvipsnames]{xcolor} 
\usepackage{setspace} 

\usepackage{longtable}
\usepackage{multirow}

\usepackage{dsfont}

\usepackage{tabularx}

\usepackage{tikz}
\usetikzlibrary{backgrounds}
\usetikzlibrary{calc}
\usetikzlibrary {arrows.meta}

\tikzset{default/.style={circle, draw, minimum size=7mm}}

\usepackage[round]{natbib}
\usepackage{amsthm}

\usepackage{mathtools}

\providecommand{\keywords}[1]
{
  \small	
  \textbf{\textit{Keywords---}} #1
}

\usepackage{mathtools} 

\usepackage{appendix}

\PassOptionsToPackage{hyphens}{url} 
\usepackage{hyperref}

\hypersetup{
    colorlinks=true,
    citecolor=Blue,
    linkcolor=Blue,
    filecolor=Plum,      
    urlcolor=Magenta,
    pdftitle={Overleaf Example},
    pdfpagemode=FullScreen,
    }

\usepackage{float}

\usepackage{subcaption}

\usepackage{booktabs}

\usepackage{todonotes}

\newcommand{\encircle}[2][]{%
  \tikz[baseline=(char.base)]{%
    \node[shape = circle, draw, inner sep = 1pt]
    (char) {\phantom{\ifblank{#1}{#2}{#1}}};%
    \node at (char.center) {\makebox[0pt][c]{#2}};}}
\robustify{\encircle}

\makeatletter
\newcommand*{\rom}[1]{\expandafter\@slowromancap\romannumeral #1@}
\makeatother 

\usepackage{arydshln}

\usepackage{authblk}
\title{Evaluating competitiveness in UEFA’s New Champions League Format}
\author[1,2]{Karel Devriesere\thanks{Corresponding author: karel.devriesere@ugent.be}}
\author[1,2]{Dries Goossens}
\author[3]{Frits Spieksma}
\affil[1]{Faculty of Economics and Business Administration, Ghent University, Tweekerkenstraat
2, 9000 Ghent, Belgium}
\affil[2]{Core lab CVAMO, FlandersMake@UGent, Belgium}
\affil[3]{ Eindhoven University of Technology, Department of Mathematics and Computer Science, The
Netherlands}
\date{}

\begin{document}
\maketitle

\abstract{Recently, UEFA changed the group stage of its international soccer competitions to an incomplete round robin tournament. Previously, teams were divided into groups, each playing a double round robin tournament with a resulting ranking table. In contrast, the new format has all teams competing in one league, producing a single ranking. We investigate the effect of the new format on the number of competitive matches in the UEFA Champions League. A match is non-competitive if the prize for at least one opponent does not depend on the match outcome, or if there exists an opportunity for both opponents to collude; otherwise, we call a match competitive. Using Monte Carlo simulations, we show that the new format results in more competitive matches than the old format.}


\section{Introduction} 

The UEFA Champions League—widely regarded as one of the most prestigious football club competitions —has recently undergone a significant change in the structure of its multi-stage format. In the previous format, the competition began by  partitioning teams into separate groups, each organized as a double round-robin tournament (i.e., every team plays the others in its group twice), with its own ranking table. In contrast, the new format begins with a so-called incomplete round-robin (iRR), in which each team plays against eight different opponents, and all teams are collectively ranked in a single unified table. A precise description of these formats is given in Section \ref{sec:reform}. According to UEFA, this revised format is intended to ``result in more competitive matches'' \citep{UEFA2024}. In this work, we examine whether UEFA's claim is justified.

A sports competition typically consists of a series of matches whose outcomes determine the allocation of specific prizes among competitors. The goal of the first stage of the UEFA Champions League is to determine the set of 16 teams that qualify for the knockout stage. In the previous format, teams could qualify for the knockout stage by finishing as one of the top two teams in their respective group. In the new format, teams can qualify in two ways for the knockout stage: either by direct qualification (awarded to the top eight teams) or by qualification throughout a knockout phase (granted to teams ranked 9th to 24th), from which additional spots in the knockout stage are subsequently earned. The latter prize is evidently less desirable. We define a match as competitive if a priori both teams have an incentive to exert full effort so as to increase their chances of securing the best possible prize. We identify three distinct types of non-competitive matches: (a) asymmetric matches, in which exactly one of the two teams is indifferent with respect to the match outcome (e.g., because it has already secured its prize); (b) stakeless matches, where the outcome has no impact on the prize allocation for either team; and (c) collusive matches, in which both teams still have something at stake, but where they can agree on a specific outcome that secures the desired prize for both. Thus, we classify the state of a match into one of these four categories: competitive, asymmetric, stakeless, or collusive (see Section~\ref{sec:prize_status} for examples). 
We emphasize that the competitiveness of a match is determined a priori, and hence does not depend on its outcome or the margin by which it was decided (e.g.\ the 2024/25 Champions League final between PSG and Inter was definitely a competitive game, even though PSG obtained an overwhelming 5-0 victory). 

Ideally, the race for the prizes remains close, culminating in an exciting climax on the final day of the competition. 
According to the outcome uncertainty hypothesis proposed by \citet{rottenberg1956baseball}, if prizes are decided before the competition concludes, the remainder of the competition will not generate much interest. Indeed, both TV viewership and stadium attendance seem to be positively linked with outcome uncertainty \citep{forrest2002outcome, garcia2002determinants, perez2017factors, cox2018spectator, schreyer2018game}. 
Fan engagement is also expected to decline significantly once their favorite team no longer has a chance of winning any of the prizes \citep{villar2009sports, pawlowski2015competition, losak2024does}. 
Moreover, a team's level of effort is likely to decline once its prize outcome is determined, compared to when it is still in contention \citep{szymanski2003economic, page2009stakes, courty2024sports}.
Finally, the integrity of the competition is another reason to avoid non-competitive matches \citep{ vanwersch2024grayness}. Collusive matches provide both teams with an incentive for match-fixing. Asymmetric and stakeless matches have also repeatedly been linked to match-fixing \citep{Duggan2002, jetter2017good, Elaad2018, Caruso2009, feddersen2023contest}. Indeed, when the cost of missing out on a prize exceeds the potential cost of corruption, the team still in the running for a prize might be tempted to try to bribe its indifferent opponent. 

The topic of non-competitive matches has been studied before for major football competitions. \citet{Guyon2020, chater2021, GuajardoKrumer2024} and \citet{stronka2024demonstration} study the 2026 FIFA World Cup which was originally designed to have a group stage with groups of three. They show that the switch from even-numbered groups to odd-numbered groups increases collusion opportunities and propose remedies such as alternative designs and dynamic scheduling. Next to collusive matches, \citet{chater2021} also estimate the sum of asymmetric and stakeless matches for the FIFA World Cup and find that the schedule in which the two strongest teams play against each other in the last round produces the lowest number of competitive games. \citet{csato2023avoid} studies the effect of tie-breaking rules on collusion opportunities in the 2022/23 UEFA Nations League and finds that competitiveness can be promoted by preferring goal difference over head-to-head results as the primary tie-breaking rule. \citet{CsatoMolontayPinter2024} compare the expected number of asymmetric and stakeless matches over the 12 possible schedules for the previous Champions League format. They find that these types of matches are minimized when the strongest team plays at home in the last round against one of the middle teams. Recently, the number of non-competitive matches has been approximated for the new Champions League format by a simple heuristic \citep{gyimesi2024competitive}. It was found that the new Champions League format reduces the number of asymmetric and stakeless games compared to the previous format.

Most of the literature considering asymmetric, stakeless and/or collusive matches has focused on tournaments with a small number of teams, where these matches can be identified with a few rules of thumb. However, for a tournament the size of the new Champions League format, determining exactly whether a team's prize is secured is computationally challenging. In this paper, we deal with this by using integer programming. Moreover, we study how to detect the opportunity for collusion for a large number of teams. To compare the number of competitive and non-competitive matches between the group stage format and the new iRR format of the UEFA Champions League, we use Monte Carlo simulation, which is one of the most popular methods to compare tournament formats \citep{devriesere2025tournament}. In this way, this paper contributes to the growing literature dealing with how tournament design impacts competitiveness.

This paper is organized as follows. Section \ref{sec:reform} discusses the reform of the Champions League format. In Section~\ref{sec:prize_status}, we define (un)competitive matches and illustrate these concepts with examples. Section~\ref{sec:methodology} discusses the methodology to perform a comparative analysis between the old and new Champions League format. Section \ref{sec:results} presents the results of our simulation study, building up to the conclusions in Section~\ref{sec:conclusion}.

\section{The UEFA Champions League reform}\label{sec:reform}

Since 2003, the first stage of the UEFA Champions League has featured a group stage comprising 32 teams partitioned into eight groups of four. Each group followed a double round-robin format, in which every team played each of the other three teams twice, once at home and once away. The top two teams from each group advanced to the knockout stage. Entry into the group stage was determined either by direct qualification - granted to e.g.\ champions of higher-ranked national associations - or by progression through one or more qualifying rounds. Beginning in 2008, these qualifying rounds were restructured into two separate tracks: one for league champions from lower-ranked associations, and another for non-champions from higher-ranked associations. In the remainder of this text, we label this ``the group stage format''.

Beginning with the 2024/25 season, UEFA implemented a new competition format for the Champions League, replacing the traditional group stage with an incomplete round-robin (iRR) structure (see e.g.\ \citet{li2025beyond} and \citet{devriesere2025redesigning}). Under this revised system, teams are no longer divided into groups but instead compete within a single league table. The number of participating clubs has been increased from 32 to 36, with the additional slots allocated as follows: one place each for the two highest-performing national associations from the previous season, one place for the third-ranked club from the fifth-best association, and one additional berth in the champions path. In this new iRR format, also known as the ``league phase'', each team plays eight matches against eight distinct opponents, without the need for transitive matchups (i.e., matches A vs.\ B, and B vs.\ C do not imply a match between A and C).

Following this phase, a two-legged play-off round is contested between teams ranked 9th to 16th (seeded) and those ranked 17th to 24th (unseeded). The winners of these fixtures join the top eight teams from the league phase in the knockout stage, after which the competition resumes its traditional knockout format, including quarter-finals, semi-finals, and the final. Notably, similar structural reforms have also been applied to UEFA's Europa League and Conference League competitions.

Closely related to the incomplete round-robin (iRR) format is the Swiss system (see e.g.\ \citet{Sauer2024swiss}). Like the iRR format, the Swiss format involves teams playing a fixed number of matches against only a subset of all possible opponents. However, the key distinction lies in the scheduling methodology: whereas the iRR format establishes the entire match schedule prior to the start of the competition, the Swiss system schedules rounds dynamically. Pairings in each round are determined based on predefined rules that take into account prior results, typically matching teams with similar performance records as the tournament progresses.

\section{Competitive matches: definitions and examples}\label{sec:prize_status}

\begin{figure}[t!]
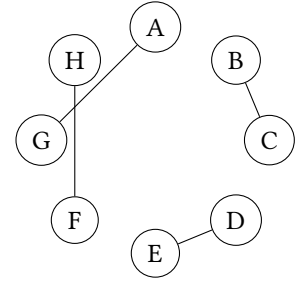

    \centering
    \caption{Example collusion opportunities}
    \begin{subfigure}[t]{0.3\textwidth}
        \caption{Ranking table 1}
        \centering
        \begin{tabular}{c|ccc}
            Rank & Team & Pts & GD \\
            \hline
            1 & A & 24 & +5 \\
            2 & B & 23 & +1 \\
            3 & C & 23 & +1 \\
            \hdashline
            4 & D & 23 & +1 \\
            5 & E & 20 & +1 \\
            6 & F & 16 & -2 \\
            7 & G & 10 & -2 \\
            \hdashline
            8 & H & 9 & -5 \\
        \end{tabular}
        \label{tab:example2}
    \end{subfigure}
    \hfill
    \begin{subfigure}[t]{0.3\textwidth}
        \caption{Ranking table 2}
        \centering
        \begin{tabular}{c|ccc}
            Rank & Team & Pts & GD \\
            \hline
            1 & A & 23 & +2 \\
            2 & B & 23 & +1 \\
            3 & C & 23 & +1 \\
            \hdashline
            4 & D & 23 & +1 \\
            5 & E & 15 & +0 \\
            6 & F & 14 & +0 \\
            7 & G & 10 & -1 \\
            \hdashline
            8 & H & 7 & -4 \\
        \end{tabular}
        \label{tab:example3}
    \end{subfigure}
    \hfill
    \begin{subfigure}[t]{0.3\textwidth}
        \caption{Final round matches and their state according to the ranking tables.}
        \centering
        \begin{tabular}{c|cc}
            Match & Table 1 & Table 2 \\
            \hline
            A-G & collusive & collusive \\
            B-C & competitive & competitive \\
            E-F & asymmetric & stakeless \\
            D-H & competitive & competitive
        \end{tabular}
        \label{fig:graph}
    \end{subfigure}
    \label{Collusion_example}   
\end{figure} 

In line with the literature (e.g.\ \citet{chater2021, CsatoMolontayPinter2024}), we distinguish three types of non-competitive matches: stakeless, asymmetric, and collusive matches. In a stakeless match, both teams are indifferent about the match outcome: no outcome can have an impact on the prize that they collect at the end of the tournament. If exactly one of the opponents is indifferent, but the other can still lose the prize it currently holds or obtain a better one depending on the match outcome (and possibly the outcome of other matches), the match is said to be asymmetric. A match is collusive if both teams have something at stake, and there exists an outcome such that both teams guarantee their best possible prizes. This does not imply that collusion necessarily occurs, but rather that both teams possess aligned incentives to facilitate such an outcome. In a competitive match, neither team is indifferent, and their objectives are incompatible, in the sense that collusion cannot help both teams achieve their best possible prizes. Hence, any game that is not asymetric, stakeless, or collusive is considered to be competitive. We emphasize that in our context the prize that a team can win depends only on its position in the final ranking.

Figure \ref{Collusion_example} illustrates each of the states a match can be in. It shows two hypothetical ranking tables of a double round robin tournament, as well as a table with the final round matches, indicating for each match its state under both ranking tables. Similar to the UEFA Champions League, teams are ranked using a point-based system where teams earn 3 points for a win, 1 point for a tie and no points for a loss; we call this a tournament with score vector (3,1,0). Goal difference (GD), i.e.\ the difference between the sum of the goals that a team scored over the matches minus the sum of the goals it conceded, is the primary tie-breaker, followed by goals scored (not shown in the tables). There are two prizes: top 3 and top 7. It can be verified that both tables are possible outcomes of a double round robin tournament where the matches indicated in Figure \ref{fig:graph} still need to be played.

Consider first the standing as in Figure~\ref{tab:example2}. The game B-C is competitive, since both teams are still in the running for top 3, but could also drop to a top 7 ranking. For example, if B and C score the same number of goals, they can be overtaken by D and E, while a win guarantees a place in the top 3. Note that although their objectives are incompatible, there is a possibility that they both reach top 3 (in case of favorable outcomes in the matches D-H and E-F). Furthermore, we see that F is eliminated from the top 3 and guaranteed to finish at least in the top 7, while E can still reach top 3. Hence, this match is asymmetric. If we look at Figure~\ref{tab:example3}, we see that the match E-F is stakeless, as both are eliminated from the top 3 but are guaranteed to finish in the top 7. The match A-G is in both cases collusive, but the outcome by which both teams guarantee their prize differs. In the ranking in Figure~\ref{tab:example2}, a loss for A with a goal difference of less than 3 ensures that both A and G guarantee their prize (top 3 and top 7). Hence, in this scenario the match A-G is collusive. If we now consider the standing as in Figure~\ref{tab:example3}, we see that a tie ensures that both A and G are guaranteed to finish in the top 3 and top 7, respectively. Hence, in this scenario the match A-G is collusive as well. In fact, in a tournament with (3,1,0)-score vector, a tie and a small loss are the only two outcomes which make a match prone to collusion. Note that in both outcomes, goal difference plays a crucial role as the primary tie-breaking rule.

We further note that there are other ways to express the competitiveness of matches. \citet{schilling1994importance}, for example, quantifies the importance of a match with respect to a prize $x$ for team $i$ as the difference between the conditional probability of reaching prize $x$ if $i$ wins and reaching the prize if $i$ loses. This measure has extensively been used to compare tournament formats (e.g.\ \citet{Scarf2009}). \citet{GoossensBelienSpieksma2012} measure match importance on a continuous scale based on the number of teams that are eliminated for certain prizes. Next, \citet{Geenens2014} and \citet{corona2017importance} quantify the decisiveness of a match based on entropy-related measures while \citet{goller2024general} consider event-importance. The important difference with the classification used in this paper is that the above-mentioned papers consider competitiveness as a probabilistic measure. For example, when there is only a marginally small chance that collusion will be profitable, the classification used here considers the match as collusive while other methods might rather consider the match as competitive. Even though these measures can be insightful, they require arbitrary cutoff points to classify matches as ``sufficiently'' competitive. In contrast, our definitions require no such assumptions, allowing a robust comparison between different formats.

In order to determine whether a match is competitive, asymmetric, stakeless or collusive, we need to establish for both teams whether they have something at stake. If teams have nothing at stake, we say that the team's prize is fixed. A team's prize is said to be fixed if (i) the team has been eliminated from contention for any higher prize and (ii) it is guaranteed to receive the prize for which it currently qualifies. Determining whether the former condition holds is known as the \emph{elimination problem} \citep{bernholt1999football}, while the verifying the latter was termed the \emph{guaranteed points placement problem} \citep{christensen2015soccer}. 

If the score vector is (1,0), meaning matches cannot end in a tie and teams earn 1 point for a win and 0 for a loss, the so-called \emph{magic number} or \emph{clinching number} indicates the number of additional wins a team needs to secure winning the championship \citep{husted2021improving}. In this case, deciding whether a team is eliminated for the first position can be solved in polynomial time \citep{schwartz1966possible, robinson1991baseball, wayne2001new, adler2002baseball, gusfield2002structure}. However, building on the results of \citet{hoffman1970team}, \citet{mccormick1996fast} prove that determining whether a team is eliminated from the top $k > 1$ positions is already NP-complete even for this simple score vector. For a tournament with a (3,1,0)-score vector, the elimination problem was proven to be NP-complete by \citet{bernholt1999football}, while the guaranteed points placement problem was proven to be coNP-complete by \citet{christensen2015soccer}, even when only considering the first and last position, respectively. However, depending on the remaining matches, algorithms exist that can solve these problems in polynomial time \citep{schlotter2018connection}. In practice, however, both problems are usually solved with integer programming \citep{ribeiro2005application, GoossensBelienSpieksma2012, raack2014standings, gotzes2022bounding}.

\section{Methodology}\label{sec:methodology}

In this section, we first show how we identify whether a match is asymmetric, stakeless, collusive or competitive (Section~\ref{sec:measuring_compet}). Next, in order to simulate the Champions League under the previous and current format, we need to generate a feasible draw (Section~\ref{sec:draw}) together with a schedule (Section~\ref{sec:schedule}). Moreover, for each match we need to simulate an outcome (Section~\ref{sec:match}). Finally, we need to make assumptions about the set of clubs, prizes and tie-breaking rules (Section~\ref{sec:selected_clubs}).

\subsection{Identifying the state of a match}\label{sec:measuring_compet}

Here, we show how to identify the state of a match. We first show how solving both the elimination problem and the guaranteed points placement problem for a given team $t$ determines whether that team has something to play for. Thus, solving these two problem for all teams reveals whether or not a match is stakeless, or asymmetric. Second, we describe a procedure to identify whether a match is collusive. Throughout this section, we assume that goal difference is the primary tie-breaking rule, in accordance with the official tie-breaking rules in the iRR format (see Section \ref{sec:selected_clubs}).

We first recall the formulation of the elimination problem proposed by \citet{ribeiro2005application}, which aims to determine the highest possible rank team $t$ can still achieve. Let $T$ be the set of teams and $M$ the set of matches that remain to be played. We define $y_i$ as a binary variable that is 1 if team $i \in T$ finishes with strictly more points than a the given team $t$. The binary variables $w_{m}, d_{m}, l_{m}$ are 1 if match $m = (i,j)$, where $i$ is the home and $j$ is the away team, results in a win, tie or loss for $i$, respectively. The sets $M_i^{h}$ and $M_i^{a}$ denote the remaining home ($h$) and away ($a$) matches of team $i$, respectively. The variable $p_i$ represents the number of points of team $i$ after all matches have been played. Let $B_i$ be the current points of team $i$. Then, we can construct the following integer program: \\

\begingroup
\begin{align}
    \text{min} \ \ z = & \sum_{i \in T: i \neq t}y_{i}\label{obj_ip1} \\
    & w_{m} + d_{m} + l_{m} = 1 & \forall \ m \in M \label{cons1_ip1}\\
    & p_{i} = B_{i} + \sum_{m \in M_i^{h}}(3w_{m}+d_{m}) + \sum_{m \in M_i^{a}}(3l_{m}+d_{m}) & \forall \ i \in T \label{cons2_ip1}\\
    & p_{i} - p_{t} \leq (B_i+3|M_i^h \cup M_i^a|-B_t)y_{i} & \forall \ i \in T: i \neq t \label{cons3_ip1} \\
    & w_m, d_m, l_m \in \{0,1\} & \forall \ m \in M \label{var:w_m}\\
    & y_i \in \{0,1\} & \forall \ i \in T: i \neq t \label{var:y_i}
\end{align}
\endgroup

  The objective function \eqref{obj_ip1} minimizes the number of teams ranked above $t$. Constraints \eqref{cons1_ip1} state that each match $m$ should result in exactly one of the three possible outcomes. Constraints \eqref{cons2_ip1} determine the points of each team after all matches have been played. Finally, constraints \eqref{cons3_ip1} ensure that $y_i$ is set to 1 if $i$ finishes with strictly more points than $t$. Thus, team $t$ is eliminated from the top $k$ positions if and only if $z$ is at least $k$. Observe that when $t$ has at least one remaining game, given that goal difference is the first tie-breaking rule in case of equal points, it is sufficient to minimize the number of teams that strictly rank above $t$. Indeed, we can set the number of goals scored by $t$ in its remaining matches arbitrary large and the goals scored of other teams as small as possible, i.e. 0-0 for a tie and 1-0 or 0-1 for a home and away win, respectively.

To model the guaranteed points placement problem, we replace constraints \eqref{cons3_ip1} with \eqref{cons4_ip1}. By maximizing $z$, we achieve the largest possible number of teams finishing with no more points than $t$. Then, team $t$ is guaranteed to finish in the top $k$ if and only if $z$ is at most $k-1$.

\begin{align}
    \text{max} \ \ z = & \sum_{i \in T: i \neq t}y_{i}\label{obj_ip2} \\ 
    & p_{t} - p_{i} \leq (B_t+3|M_t^h \cup M_t^a|-B_i)(1-y_{i}) & \forall i \in T: i \neq t \label{cons4_ip1} \\
    & \eqref{cons1_ip1}, \eqref{cons2_ip1}, \eqref{var:w_m}, \eqref{var:y_i} \nonumber
\end{align}

Observe that when $t$ has at least one remaining game, given that goal difference is the first tie-breaker in case of equal points, it is sufficient to maximize the number of teams that finish with the same number of points as $t$. Indeed, we can assume $t$ loses its remaining games with an arbitrary large goal difference such that any other team that finishes with an equal number of points as $t$ ranks above $t$. Solving both problems described above determines whether teams still have something at stake or not. This information is sufficient to identify asymmetric and stakeless matches. 

If a match is neither asymmetric nor stakeless, it may yet be collusive. To detect whether a match $(i,j)$ is prone to collusion we distinguish two cases: when collusion is realized through a tie or a small loss. A further distinction can be made based on how many games are left. In order to identify whether a match is collusive through a tie, we solve the guaranteed points placement problem for both $i$ and $j$, assuming the match ends in a tie. If after this match, $i$ and $j$ still have at least one other match (against different opponents), we do not have to take goal difference into account since both $i$ and $j$ may lose their last game with an arbitrary large goal difference. In case $(i,j)$ is the last game of $i$ and $j$, we cannot neglect goal difference, as there can be teams who can finish with the same number of points as $i$ and/or $j$ by scoring the same number of goals, but with a worse goal difference. In case all teams have at most one remaining game left, we can simply iterate over all remaining matches to determine the maximum number of teams that can still finish above $i$ ($j$). We first check, for each match other than $(i,j)$, whether both teams in that match can rank above $i$ ($j$), taking into account the tie-breaking rules. If not, we check whether one of the teams finishes with at the least the same number of points as $i$ ($j$), given it wins that match. This way, we count the maximum number of teams that can finish above $i$, and we do the same for $j$. Finally, note that a collusion by a small loss is only possible for match $(i,j)$ if this match is the final match for the losing team. First, we determine whether $i$ or $j$ guarantees its prize when losing with only one goal conceded, while the other team guarantees its prize with a high number of goals scored. For practical purposes, we consider small losses up to a difference of five goals. This is in line with \citet{chater2021}, who check all possible goal differences between -5 and +5 to check collusion opportunities. If the answer is yes, we check whether one of the outcomes $(1,0),(2,0),(3,0),(4,0),(5,0)$, assuming here w.l.o.g. that $i$ is the team that must win, guarantees the best possible prize for both teams. If yes, the match is collusive by small loss.

\subsection{Draw}\label{sec:draw}

\subsubsection{Group stage format}

In the group stage format, 32 teams are drawn in eight groups of four teams each. Since the aim of the groups is to balance teams of similar strength over the groups, the 32 teams are first partitioned into 4 pots. Since the 2018/19 season, pot 1 contains the former Champions and Europa League winners, together with the champions of the six (or seven) countries with the highest UEFA association coefficients. Pots 2, 3 and 4 contain the other teams, seeded based on their UEFA association coefficients. Each group consists of one team from each pot.

The draw procedure works as follows. First, clubs from pot 1 are randomly assigned to groups. Next, clubs from pot 2 are drawn. For each drawn team, \emph{“the computer indicates which groups are available for this club, and a bowl is prepared containing balls representing each of the groups into which the club could be drawn''}\citep{UEFA2023draw_pairings}. This process is repeated for clubs of pot 3 and finally for clubs of pot 4. In order to simulate the group stage draw, we propose to randomly assign teams to groups, in increasing order of the pots, and to use an integer program to check if the draw can be completed in case the team is assigned to its drawn group. If not, we remove the group from the team's available groups and draw a new group, until we have found a group such that the draw can be completed. For this purpose, let $T$ be the set of teams, $P_h$ be the teams of pot $h$ (with $|P|$ pots in total) and $C$ be the set of associations that are represented in the Champions League. Let $T_c$ be the teams of association $c$ and $a(i)$ be the association of team $i \in T$. Let $G$ be the set of groups. We use a binary variable $x_{ig}, \forall i \in T, \ g \in G$ that is 1 if team $i$ is assigned to group $g$ and 0 otherwise. Let $\psi_{ig}$ be a parameter that is 1 if $i$ was allocated to group $g$ in previous iterations, and 0 otherwise.

\begin{align}
    & \sum_{i \in T_c}x_{ig} \leq 1 & \forall \ c \in C, \ \forall \ g \in G \label{draw_G:c1} \\
    & \sum_{g \in G} x_{ig} = 1 & \forall \ i \in T \label{draw_G:c2} \\
    & \sum_{i \in P_h}x_{ig} = 1 & \forall \ g \in G, \ \forall \ h \in 1, ..., |P| \label{draw_G:c3} \\
    & x_{ig} \geq \psi_{ig} & \forall \ i \in T, \ \forall g \in G \label{draw_G:c4} \\
    & x_{ig} \in \{0,1\} & \forall \ i \in T, \ \forall \ g \in G \label{draw_G:x} 
\end{align}

Constraint \eqref{draw_G:c1} guarantees that each group contains at most one team from association $c \in C$. Constraint \eqref{draw_G:c2} assigns each team to exactly one group, while constraint \eqref{draw_G:c3} ensures that each group consists of exactly one team from each pot. Previously drawn matches are fixed by \eqref{draw_G:c4}.

\subsubsection{iRR format}

For the draw of the iRR format, the following constraints need to be satisfied \citep{UEFA2024draw}: 

\begin{enumerate}
    \item Each team faces exactly two different opponents from each pot, one at home and one away
    \item A team is never matched with a team from its domestic association
    \item A team faces at most two opponents from the same association
\end{enumerate}

With the expansion to 36 teams, by default up to five teams of the same association are now allowed to participate. In addition to matching teams, also the home and away status is determined. Similarly to previous years, the draw is conducted iteratively. First, a team from pot 1 is drawn. Next, its eight opponents, two from each pot, are drawn, indicating for each opponent whether it is faced at home or away. This procedure is repeated for all remaining teams from pot 1. Next, a team from pot 2 is drawn, together with its 6 opponents from pots 2-4. This process is repeated for all teams from pot 2, next for pot 3 and finally for pot 4.

Similarly to the group stage, we now present an integer program that checks whether the draw can be completed if a match is drawn. We use a binary variable $m_{ij}, \forall i, j \in T$ that is 1 if $i$ plays home against $j$ and 0 otherwise. Let $\omega_{ij}$ be a parameter that is 1 if $i$ and $j$ were matched in a previous iteration, and 0 otherwise. Before committing to opponent $j$ being drawn for team $i$, we check if the following set of constraints yield a feasible solution:

\begin{align}
    & \sum_{j \in T_{a(i)}}(m_{ij}+m_{ji}) = 0 & \forall i \in T \label{draw:c8} \\
    & \sum_{j \in T_c}(m_{ij} + m_{ji}) \leq 2 & \forall \ i \in T, \forall \ c \in C: a(i) \neq c \label{draw:c1} \\
    & \sum_{j \in P_h}m_{ij} = 1 & \forall \ i \in T, \forall \ h = 1,\dots,|P| \label{draw:c3} \\
    & \sum_{j \in P_h}m_{ji} = 1 & \forall \ i \in T, \forall \ h = 1,\dots,|P| \label{draw:c4} \\
     & m_{ij} + m_{ji} \leq 1 & \forall \ i \in T \label{draw:c2} \\
    & m_{ij} \geq \omega_{ij} & \forall \ i, j \in T \label{draw:c5}\\
    & m_{ii} = 0 & \forall \ i \in T \label{draw:c6} \\
    & m_{ij} \in \{0, 1\} & \forall \ i, j \in T \label{draw:c7}
\end{align}

Constraint \eqref{draw:c8} ensures that a team does not play against a team from the same association. Constraint \eqref{draw:c1} ensures that each team is matched with at most two teams from the same (but different than its own) association. Constraints \eqref{draw:c3} and \eqref{draw:c4} guarantee that a team faces exactly one team from each pot at home and one team from each pot away. Constraint \eqref{draw:c2} states that a team is not matched to the same opponent twice. Previously drawn matches are fixed by \eqref{draw:c5}. Naturally, a team does not face itself, which is stated by \eqref{draw:c6}. 
If the integer program is feasible, we pick the drawn team, update the value for the corresponding $\omega$ parameter, and continue the draw. If not, then we remove team $j$ from the available opponents of $i$ and draw a new team. This approach mimics the official iterative procedure of UEFA, and guarantees that we will find a feasible draw (if one exists).

Constraints \eqref{draw:c1}-\eqref{draw:c7} do not avoid the possibility of a draw that needs 9 matchdays to be completed. Such a scenario is possible, as is shown in \citet{Guyon2025}. Since this only has an extremely small probability, however, we do not enforce this in the model presented above. Instead, if such a scenario arises, which would be detected by the integer program in the next section, we simply perform another draw.

\subsection{Schedule}\label{sec:schedule}

Given a draw, a feasible schedule needs to be constructed. We now introduce the following notation. Let $B=\{1,\dots,8\}$ be the set containing the indices of the rounds, and $S=\{1,\dots,|S|\}$ the set containing the indices of the time slots. Let $S$ be partitioned into subsets $S_b$ containing the indices of time slots of round $b \in B$. In each time slot $s \in S$ a certain number of matches $m_s$ need to be played. Each team plays exactly one match in each round, in one of the available time slots. For example, the second round of the 2024/25 competition was played on Tuesday and Wednesday of the first week of October. Both Tuesday and Wednesday contain two time slots: 18:45 pm and 21:00 pm. In the early time slot, two matches need to be planned while in the late time slot seven matches need to be planned, for a total of 18 matches in that round. It is important in this work to consider the specific time slots, and not only the rounds, because match outcomes of earlier slots in same round potentially impact the prize status of teams in the remaining matches. 

For both the group stage and the iRR format, UEFA gives surprisingly little information on how the matches are scheduled. The only additional scheduling requirement specific to the iRR format, beyond those implied by the published draw rules, is as follows: “A club does not play more than two home or two away matches in a row and each club plays one home match and one away match across the first two matchdays and across the last two matchdays.” \citep[Article 17]{UEFA2024schedule}. The same requirement was also present in the previous format \citep{CsatoMolontayPinter2024}. 

\subsubsection{Group stage format}

\citet{CsatoMolontayPinter2024} observe that from the 2021/22 season onwards, rounds 4/5/6 are the mirror image of rounds 3/2/1. In the absence of other requirements, they list 12 valid schedules, characterized by the matches in the last two rounds. These schedules are listed in Table \ref{tab:schedules_groups}. The name of the schedule reveals the matches of the team from pot 1, e.g. in schedule 1231, the team from pot 1 plays a home game against the team from pot 2 in round 5 and plays away against the opponent from pot 3 in round 6. Rounds can have up to 4 time slots, 2 on Tuesdays and 2 on Wednesdays, except for the last round where all matches of the same group are played at the same time. For round 5, in some groups one match was played at 18:45 and the other at 21:00, while in other groups both matches were played at 21:00. Since it is unclear why in some cases matches are played in separate time slots, we make the conservative assumption that in every round, all groups play their matches simultaneously. Hence, in the group stage the number of time slots coincide with the number of rounds. Note that this leads to a small underestimation of the number of non-competitive matches. 

\begin{table}[!ht]
    \centering
    \caption{All 12 valid schedules for the UEFA Champions League group stage}
    \begin{tabular}{c|cc}
    Schedule & Round 5 & Round 6 \\
    \hline
    1231 & (1,2) & (3,1) \\
        & (4,3) & (2,4) \\
    2113 & (2,1) & (1,3) \\ 
         & (3,4) & (4,2) \\
    1241 & (1,2) & (4,1) \\ 
        & (3,4) & (2,3) \\
    2114 & (2,1) & (1,4) \\ 
        & (4,3) & (3,2) \\
    1321 & (1,3) & (2,1) \\ 
        & (4,2) & (3,4) \\
    3112 & (3,1) & (1,2) \\ 
        & (2,4) & (4,3) 
    \end{tabular}
    \quad
    \begin{tabular}{c|cc}
    Schedule & Round 5 & Round 6 \\
    \hline
    1341 & (1, 3) & (4, 1) \\
        & (2, 4) & (3, 2) \\
    3114 & (3, 1) & (1, 4) \\
        & (4, 2) & (2, 3) \\
    1421 & (1, 4) & (2, 1) \\
        & (3, 2) & (4, 3) \\
    4112 & (4, 1) & (1, 2) \\
        & (2, 3) & (3, 4) \\
    1431 & (1, 4) & (3, 1) \\
        & (2, 3) & (4, 2) \\
    4113 & (4, 1) & (1, 3) \\
        & (3, 2) & (2, 4)
    \end{tabular}
    \label{tab:schedules_groups}
\end{table}

In general, \citet{CsatoMolontayPinter2024} found that unattractive matches occur less frequently in schedules in which, in the last round, the strongest team plays at home against one of the two middle teams compared to other schedules. To asses the effect of the schedule on the group stage format, we iterate over all the options and schedule each group according to that option. Since in practice groups may be scheduled according to different options, we also include the option where each group is scheduled according to a randomly selected, hence possibly different, schedule.

\subsubsection{iRR format}
 
Finding a feasible schedule for the iRR format is less straightforward but can be done by integer programming. Let $T$ again be the set of teams, $B$ be the set of rounds, $S_b$ be the set of time slots in round $b \in B$ and $m_s$ be required number of matches in slots $s \in S_b$, $b \in B$. Let $N_i^H$ and $N_i^A$ be the sets of drawn opponents where team $i$ plays home and away against, respectively. Let $x_{ijs}$ be 1 if match $(i,j)$ is scheduled in slot $s$, and 0 otherwise.
A feasible schedule for the UEFA Champions League satisfies the following integer program:

\begin{align}
    & \sum_{s \in S}x_{ijs} = 1 &  \forall i \in T, \ \forall \ j \in N_i^H \label{cons_irr1_h} \\
    & \sum_{s \in S}x_{jis} = 1 &  \forall i \in T, \ \forall \ j \in N_i^A \label{cons_irr1_a} \\
    & \sum_{s \in S_b}(\sum_{j \in N_i^H}x_{ijs}+\sum_{j \in N_i^A} x_{jis}) = 1 & \forall \ i \in T, \ \forall \ b \in B \label{cons_irr2} \\
    & \sum_{i \in T}(\sum_{j \in N_i^H}x_{ijs} + \sum_{j \in N_i^A}x_{jis}) = m_{s} & \forall \ s \in S \label{cons_no_matches} \\
    & \sum_{j \in N_i^H}(\sum_{s \in B_{b}}x_{ijs} + \sum_{s \in B_{b+1}}x_{ijs}) \leq 1 & \forall \ i \in T, \ \forall \ b \in \{1, |B|-1\} \label{cons_breakH_start} \\
    & \sum_{j \in N_i^A}(\sum_{s \in S_b}x_{jis} 
     + \sum_{s \in S_{b+1}}x_{jis}) \leq 1 & \forall i \ \in T, \forall \ b \in \{1, |B|-1\} \label{cons_breakA_start} \\
    & \sum_{j \in N_i^H}(\sum_{s \in S_b}x_{ijs} + \sum_{s \in S_{b+1}}x_{ijs} + \sum_{s \in S_{b+2}}x_{ijs}) \leq 2 & \forall \ i \in T, \ \forall \ b \in \{2,\dots,|B|-3\} \label{cons_3h} \\
    & \sum_{j \in N_i^A}(\sum_{s \in S_b}x_{jis} + \sum_{s \in S_{b+1}}x_{jis} + \sum_{s \in S_{b+2}}x_{jis}) \leq 2 & \forall \ i \in T, \ \forall \ b \in \{2,\dots,|B|-3\} \label{cons_3a} \\
    & x_{ijs} \in \{0, 1\} & \forall \ i,j \in T, \ \forall \ s \in S\label{x_var_A} 
\end{align}

Constraints \eqref{cons_irr1_h} and \eqref{cons_irr1_a} ensure that each home and away opponent is seen exactly once over all time slots. Constraint \eqref{cons_irr2} guarantees that each team plays exactly one game in each round, while constraint \eqref{cons_no_matches} guarantees that in each time slot, the required number of matches is played. Next, constraints \eqref{cons_breakH_start} and \eqref{cons_breakA_start} forbid consecutive home or away matches in the first and last two rounds. Finally, constraints \eqref{cons_3h} and \eqref{cons_3a} enforce that teams play at most two consecutive home or away games. We remark that a similar integer program is also proposed in \citet{melkonian2024integer}.

We now describe several possible objectives for the schedule. From a commercial perspective, an important objective of UEFA is to maximize its television audience. Hence, we extend the schedule by taking into account the number of teams of the same association playing in the same time slot, as well as the number of matches between teams from pot 1 that are scheduled in the same time slot. Ideally, these matches should be spread out as much as possible. Therefore, we create a binary variable $y_{cs}$ that is 1 if a team from association $c$ plays in slot $s$ and a binary variable $z_s$ that is 1 if a team from pot 1 plays in slot $s$.


\begin{align}
     \mathcal{M}_1: \nonumber \\
    \text{max} \quad & \ \sum_{s \in S}(\sum_{c \in C}y_{cs} + z_s) \label{obj=y+Z}\\
    & y_{cs} \leq \sum_{i \in T: a(i) = c}(\sum_{j \in N_i^H}x_{ijs} + \sum_{j \in N_i^A}x_{jis}) & \forall \ c \in C, \ \forall \ s \in S \label{cons_y} \\
    & z_s \leq \sum_{i \in P_1}(\sum_{j \in N_i^H}x_{ijs} + \sum_{j \in N_i^A}x_{jis}) & \ \forall \ s \in S \label{cons_z}\\
    & y_{cs} \in \{0,1\} & \ \forall \ c \in C, \forall \ s \in S, \label{var_y} \\
    & z_{s} \in \{0,1\} & \ \forall \ s \in S, \label{var_z} \\
    & \eqref{cons_irr1_h} - \eqref{x_var_A} \nonumber
\end{align}

Constraint \eqref{cons_y} regulates that $y_{cs}$ can only be 1 if at least one team from association $c$ plays a game in slot $s$. Similarly, constraint \eqref{cons_z} ensures that $z_s$ can only be 1 if at least one team from pot 1 plays a game in slot $s$. Combining \eqref{cons_irr1_h} - \eqref{x_var_A} with \eqref{obj=y+Z}-\eqref{var_z} gives the formulation $\mathcal{M}_1$.

From a competitiveness perspective, it is worthwhile to compare schedules in which matches of teams with similar strength are scheduled as much as possible at either the beginning or end of the scheduling period. This can be done, for example, by making an ordinal ranking of the teams based on some criterion such as Elo rating. Let $r_i$ be the position of team $i \in T$ in this ranking. Let $b_{|B|-2}^l$ be the index of the last time slot in round $|B|-2$. Then, the formulation $\mathcal{M}_2$ pushes matches of teams with similar strength to the later time slots in the last two rounds.


\begin{align}
    \mathcal{M}_2: \nonumber \\
    \text{min} \quad & \sum_{b \in \{|B|-1,|B|\}}\sum_{s \in S_b} (\sum_{j \in N_i^H}|r_i - r_j|^{(s - b_{|B|-2}^l)}x_{ijs} + \sum_{j \in N_i^A}|r_i - r_j|^{(s - b_{|B|-2}^l)}x_{jis}) \\
    & \eqref{cons_irr1_h} - \eqref{x_var_A} \nonumber
\end{align}

Let $b_3^f$ be the index of the first time slot of the third round. Then, formulation $\mathcal{M}_3$ pushes matches of teams of similar strength to the earlier time slots in the first two rounds.

\begin{align}
    \mathcal{M}_3: \nonumber \\
    \text{min} \quad & \sum_{b \in \{1,2\}}\sum_{s \in S_b}(\sum_{j \in N_i^H}|r_i - r_j|^{(b_3^f - s)}x_{ijs} + \sum_{j \in N_i^A}|r_i - r_j|^{(b_3^f - s)}x_{jis}) \\
    & \eqref{cons_irr1_h} - \eqref{x_var_A} \nonumber
\end{align}


As a final type of schedule, instead of pushing matches between teams of similar strength all to the beginning or end of the scheduling period, formulation $\mathcal{M}_4$ has an additional constraint that guarantees that teams do not start or end with two games against teams from pot 1 and 2. Similarly, it also ensures that teams do not start or end with two games against teams from pot 3 and 4. The goal here is to ensure some degree of balance at the start and finish of the competition.

\begin{align}
    & \quad \mathcal{M}_4: \nonumber \\
    & \sum_{p \in P'}\sum_{b \in B'}\sum_{s \in S_b}(\sum_{j \in N_i^H \cap P_p}x_{ijs}+\sum_{j \in N_i^A \cap P_p}x_{jis}) \leq 1 & \forall i \in T, \ \forall P' \in \{\{1,2\},\{3,4\}\}, \ \forall B' \in \{\{1,2\},\{|B|-1,|B|\}\}\label{cons_balance1}\\
    & \eqref{cons_irr1_h} - \eqref{x_var_A} \nonumber
\end{align}

The main features of the schedules are summarized in Table \ref{tab:summary_new_schedules}.

\begin{table}[!ht]
    \centering
    \caption{Summary of schedules in the iRR format}
    \begin{tabular}{c|p{12cm}}
        \textbf{Schedule} & \textbf{Main characteristic} \\
        \hline
        $\mathcal{M}_0$ & No objective \\
        $\mathcal{M}_1$ & Spread (a) matches of teams from pot 1 and (b) matches of teams from the same association \\
        $\mathcal{M}_2$ & Schedule matches between teams of similar Elo strength to the final matchdays \\
        $\mathcal{M}_3$ & Schedule matches between teams of similar Elo strength to the first matchdays \\
        $\mathcal{M}_4$ & Teams do not start or end with two games against (a) two teams from pot 1 or 2 and (b) two teams from pot 3 or 4 \\
    \end{tabular}
    \label{tab:summary_new_schedules}
\end{table}

\subsection{Simulating match outcomes}\label{sec:match}

One of the most used statistical models for simulating football results is based on the Poisson distribution \citep{ley2019ranking, maher1982}. The probability that in match $(i, j)$ the home team $i$ scores $k$ goals against away team $j$ is given by:

\begin{equation}
    \mathbb{P}_{h}(k) = \frac{(\lambda_{ij}^{h})^{k}\text{exp}(-\lambda_{ij}^{h})}{k!},
\end{equation}

\noindent where $\lambda_{ij}^{h}$ denotes the expected number of goals scored by $i$ as the home team in match $(i,j)$.
Similarly, the probability $\mathbb{P}_{a}(k)$ of the away team $j$ scoring $k$ goals against $i$ can be found by substituting $\lambda_{ij}^{h}$ for $\lambda_{ij}^{a}$, the expected number of goals scored by the away team. Note that this model assumes that the results of matches are independent. 

It is common to model the $\lambda_{ij}^{h}$ for $\lambda_{ij}^{a}$ parameters as a function of the difference in strength between the two teams. Using a polynomial regression, \citet{gyimesi2024competitive} estimate the value of these lambdas as a function of the winning probability $w_{i,j}$ of team $i$ in a home game against team $j$, based on their Elo-ratings $Elo_i$ and $Elo_j$, and given as
\begin{equation}
    w_{ij} = \frac{1}{1+10^{-(Elo_i + 100 - Elo_j)/400}} \label{w_ij}.
\end{equation}
Others (e.g.\ \citet{csato2022quantifying}, \citet{stronka2024demonstration}) directly adopt the quartic polynomial estimated by \citet{footballrankings2020}. \citet{corona2019bayesian} and \citet{DagaevRudyak2019} use a generalized linear model (GLM) with a log-link function, using the UEFA club coefficients as a strength measure. 
Although more advanced methods exist, since the goal here is to compare tournament formats rather than to predict individual matches, we stick to the more common models.

We employ Elo ratings to measure team strength, as they are extensively used in the literature and have been shown to outperform UEFA club coefficients in predicting UEFA Champions League matches \citep{Csato2024elo}. For each season, we use the Elo rating of teams at the start of their first match in the Champions League of that season. Similar to \citet{gyimesi2024competitive}, the ratings from the website \url{elofootball.com} are used. In order to derive a reasonable model, we estimate the parameters by maximum likelihood, investigate the effect of the degree of the polynomial regression and compare it to a generalized linear model. Data from the Champions League group stage in seasons 2005/06 till 2022/23 were used to estimate the model parameters, resulting in a training set of 1968 matches, while seasons 2023/24 and 2024/25 were used to validate the models. It should be noted that data from the 2019/20, 2020/21, and 2021/22 seasons may be biased because several matches were played behind closed doors. 

In order to compare the performance of the models, we compute the percentage of correctly predicted match outcomes for each of the trained models, for both the group stage matches of the season 2023/24 and the league phase matches of the season 2024/25. Next to the fitted models, we also computed the historical percentage of wins (46.76\%), ties (22.86\%) and losses (30.38\%) and used those probabilities to guess the match outcomes.

Moreover, we also compute the average Rank Probability Score (RPS) of individual matches, as defined in \citet{constantinou2012solving}: 

\begin{equation}
    \text{RPS} = \frac{1}{r-1}\sum_{i=1}^{r-1}(\sum_{m=1}^{i}(p_m-e_m))^2, \label{rps}
\end{equation}

\noindent where parameter $r$ is the number of potential outcomes and $p_m$ and $e_m$ are the predicted and observed outcomes of match $m$, respectively. Intuitively, it measures how closely the predicted match outcome is to the observed outcome and is argued by \citet{constantinou2012solving} to be a better football performance metric than other prevalent methods like the log loss and Brier score. Indeed, it has been used in many other works that compare football prediction models, see for example \citet{groll2019hybrid}, \citet{ley2019ranking} and \citet{hubavcek2022forty}. A lower RPS value denotes a more accurate fit. The results are shown in Table \ref{tab:model_performance}.

\begin{table}[!ht]
    \centering
    \caption{Comparison of predictive models across two seasons. W = Win, T = Tie, L = Loss, RPS = Ranked Probability Score.}
    \renewcommand{\arraystretch}{1.2}
    \begin{tabular}{c|cccccc}
        Model & W (\%) & T (\%) & L (\%) & Correct (\%) & RPS & Season \\
        \hline
        Random guess & 46.76 & 22.86 & 30.38 & 31.25 & 23.40 & 2023/24 \\
        Cubic polynomial      & 43.75 & 30.21 & 26.04 & 59.38 & 17.33 & 2023/24 \\
        Quartic polynomial      & 44.79 & 28.12 & 27.08 & 59.38 & 17.34 & 2023/24 \\
        Quintic polynomial      & 47.92 & 25.00 & 27.08 & 59.38 & 17.34 & 2023/24 \\
        GLM          & 48.96 & 27.08 & 23.96 & 58.33 & 17.54 & 2023/24 \\
        Actual       & 46.88 & 20.83 & 32.29 & --    & --    & 2023/24 \\
        \hline
        Random guess & 46.76 & 22.86 & 30.38 & 37.50 & 23.96 & 2024/25 \\
        Cubic polynomial     & 46.53 & 20.83 & 32.64 & 54.17 & 19.12 & 2024/25 \\
        Quartic polynomial     & 47.22 & 18.06 & 34.72 & 56.25 & 19.13 & 2024/25 \\
        Quintic polynomial      & 47.92 & 16.67 & 35.42 & 56.94 & 19.19 & 2024/25 \\
        GLM          & 50.00 & 18.75 & 31.25 & 58.33 & 19.23 & 2024/25 \\
        Actual       & 53.47 & 12.50 & 34.03 & --    & --    & 2024/25 
    \end{tabular}
    \label{tab:model_performance}
\end{table}

For each model, we report the percentage of matches resulting in a home win (W), home loss (L) and tied (T). As can be seen from the model, the Poisson-based models clearly outperform a random guess based on the historical average. There seems to be little difference, however, between the performance of the Poisson-based models. 
Moreover, the models appear to perform better on the 2023/24 data, which is unsurprising given that they were trained exclusively on data from group-stage matches rather than league-stage matches. Nevertheless, the RPS values closely resemble the values of those found in the literature (see e.g. \citet{ley2019ranking}). After performing a Nemenyi test for both RPS and the percentage of correct guesses, taking the two seasons together, the 4 Poisson-based models where not found to significantly differ (the lowest $p$-value observed for RPS was 0.34 and for the percentage of correct guesses it was even 0.99). Still, because of the higher percentage correctly predicted matches in the 2024/25 league phase, we opted to use the GLM model in our computational experiments. The corresponding formulas of $\lambda_{i,j}^{h}$ and $\lambda_{i,j}^{a}$ are given in equations \eqref{eq:lambda_h} and \eqref{eq:lambda_a}, respectively. 

\begin{align}
    & \text{log}(\lambda_{i,j}^{h}) = -0.8816 + 0.0023 \times \text{Elo}_i - 0.0018 \times \text{Elo}_j + 0.1484 \label{eq:lambda_h} \\
    & \text{log}(\lambda_{i,j}^{a}) = -0.8816 + 0.0023 \times \text{Elo}_j - 0.0018 \times \text{Elo}_i \label{eq:lambda_a} 
\end{align}

\subsection{Experimental design}\label{sec:selected_clubs}

In order to adequately asses the impact of the format change, we compare the two formats on the last six Champions League seasons. Taking multiple seasons into account reduces the bias caused by the Elo distribution of the competing teams in one particular season. However, since the iRR format features 36 instead of 32 teams as in the group stage format, we need to select four additional teams. The official UEFA rules allocate one extra spot for the 5th-ranked national association, one extra spot for a team from the so called champions qualifying path and two additional spots for the best-performing associations in the previous season. For each season played according to the old format, we follow these rules in the following way: \\
\begin{itemize}
    \item[$\blacksquare$] We add the best team (according to the domestic league) that was not yet present in the original 32 teams for the two associations with the highest UEFA association coefficient of the previous season. 
    \item[$\blacksquare$] If the club that would have qualified as the 5th-ranked national association did already qualify, the additional spot is allocated to the best performing non-qualified team from the non-champions path. 
    \item[$\blacksquare$] For the additional team coming from the champions path, we select the best performing non-qualified team. Specifically, we selected the team that lost with the smallest goal difference, breaking ties randomly. If this best performing team was already selected because its association was among the two highest ranked in the previous season, we select the next best performing team.
\end{itemize}
We note that these rules resulted in a feasible draw for the above mentioned seasons, i.e. no team from the added paths resulted in more than five teams from the same association and no five teams of the same association were present in the same pot of nine teams. To construct the pots for the iRR format, the winner of the Champions League's past season was placed in pot 1, while the other teams were allocated to pots based on their club coefficients of the past season. For the season 2024/25, we delete four teams when simulating the group stage format according to a similar interpretation of these rules. 

After simulating match results, teams are ranked in both formats according to their points. However, the tie-breaking rules changed with the iRR format. In the seasons before 2024/25, ties were first broken based on head-to-head results. In contrast, in the iRR format, total goal difference is the first tie-breaking rule. It is known that tie-breaking rules affect the probability of asymmetric, collusive, and stakeless games as demonstrated in \citet{csato2023avoid} and \citet{csato2025head}. Moreover, recall that it is sufficient to use the integer programs in section \ref{sec:measuring_compet} to compute the prize status if the first tie-breaking rule is goal difference. If the first tie-breaking rule is head-to-head, additional checks are possibly needed as we cannot assume anymore that teams can score or concede an infinite number of goals. Since we want to distinguish the effect of the tie-breaking rules with that of the format, when simulating the group stage we rank teams according to the tie-breaking rules that are used in the iRR format.

Finally, we consider the following prizes:\\

\begin{itemize}
     \item[$\blacksquare$] Group stage: Top 2 (qualified), 3rd (qualified for the Europa League), 4th (eliminated)
    \item[$\blacksquare$] iRR: Top 8 (qualified for the knockout stage), 9–24 (qualified for the
    play-offs that determine qualification for the knockout stage), 25–36 (eliminated)
\end{itemize}

It can be argued that the first and second place in the group stage are different prizes, since the first place is seeded while the second place is unseeded in the knockout stage. However, in contrast to the prizes defined above, seeding does not directly lead to elimination from the tournament. Moreover, although in theory the seeding incentivizes strong performance by seeding higher ranked with lower ranked teams, in practice teams often have their own preferences against which teams they want to play. Indeed, there exists numerous examples where teams deliberately lost to obtain a lower seed so as to play against a preferred opponent \citep{devriesere2025manipulation}. Next, accounting for these alternative prizes would require reporting twice as many results, which would considerably increase the length of the paper and potentially reduce its readability. Therefore, we chose to take into account only the set of prizes mentioned above. 

\section{Results}\label{sec:results}


For both the group stage and the iRR format, we perform 10,000 simulations for each type of schedule and each season. For the group stage format, we also include the option that each group is scheduled randomly according to each of the twelve possible schedules. Hence, the total number of simulations is 13 $\times$ 6 $\times$ 10,000 = 780,000 for the group stage and 5 $\times$ 6 $\times$ 10,000 = 300,000 for the iRR format. In each simulation, we simulate the draw, which designates the set of matches that need to be played. Next, we sample match outcomes for each of these matches. Then, we use those match outcomes to evaluate competitiveness in the different types of schedules. 

\subsection{Comparison group stage and iRR format}

\begin{table}[!ht]
    \centering
    \caption{Observed percentage of each type of match, averaged over the seasons, for each format with random schedule. This corresponds to $\mathcal{M}_0$ for the iRR format and each group being randomly assigned one of the 12 possible schedules for the group stage format.}
    \begin{tabular}{c|cccc}
         Format & Stakeless (\%) & Asymmetric (\%) & Collusive (\%) & Competitive (\%) \\
        \hline
        iRR & 0.75 & 5.14 & 0.17 & 93.94 \\
        Group stage & 2.78 & 10.28 & 0.11 & 86.82 \\
        $\Delta$ & -2.03 & -5.14 & +0.06 & +7.12 \\
        \hline
        U (p) & $2.8 \times 10^9$ (< 0.01) & $2.7 \times 10^9$ (< 0.01) & $1.6 \times 10^9$ (< 0.01) & $5.4 \times 10^8$ (< 0.01)\\
        $r_{\text{rb}}$ & -0.55 & -0.49 & +0.10 & +0.70
    \end{tabular}
    \label{tab:old_vs_new}
\end{table}

To compare both formats, we compare the results when a random schedule is used. Table \ref{tab:old_vs_new} shows the average percentage of stakeless, asymmetric, collusive and the total number of non-competitive matches, respectively, for both formats. The difference between the iRR and the group stage format is shown in the row $\Delta$. 

A key observation is that the number of stakeless and asymmetric matches is much lower in the iRR format (which is in line with the results of \citet{gyimesi2024competitive}), while the percentage of collusive matches is slightly higher. To test whether these values significantly differ, we perform a Mann-Whitney U test, which is a nonparametric test to evaluate whether two independent samples follow the same distribution. In this case, we test whether the values for stakeless and asymmetric matches are higher in the group stage compared to the iRR format. Similarly, we test whether the values for collusive and competitive matches are lower in the group stage compared to the iRR format. The resulting U-statistics (U) and corresponding p-values (p) are shown in Table \ref{tab:old_vs_new}. All values were found to significantly differ from each other (p < 0.01). Next, we compute the Rank-Biserial Correlation ($r_{\text{rb}}$) to measure the strength of the association between the formats and the observed number of non-competitive matches. A value of 0 indicates no correlation, while a value of +1 and -1 indicate a very strong and weak correlation, respectively. For example, we observe a strong positive effect (+0.70) of the iRR format on competitive matches. Based on Table \ref{tab:old_vs_new}, we conclude that the iRR format performs better with respect to stakeless, asymmetric and competitive matches, while it seems to provoke slightly (but significantly) more collusive games.

We now take a closer look at the distribution of the non-competitive matches over the different time slots for the group stage format (Table \ref{tab:non_comp_matches}) and the iRR format (Table \ref{tab:non_comp_matches_irr}). Recall that for the group stage format we made the conservative assumption that the four teams from each group play their matches at the same time. Hence, the time slots coincide with the rounds and the number of matches in each time slot is 16. For the iRR format, round 7 consists of 4 time slots, while round 8 consists of a single time slot. As expected, non-competitive matches occur more frequently in later rounds. 

\begin{table}[!ht]
    \centering
    \caption{Overview of each type of match match per round/slot in the group stage format}
    \begin{tabular}{c|cccccc}
        Round & Slot & \#Matches & Stakeless (\%) & Asymmetric (\%) & Collusive (\%) & Competitive (\%) \\
        \hline
        5 & 1 & 16 & 0.91 & 17.02 & 0.3 & 81.77 \\
        6 & 1 & 16 & 15.79 & 44.67 & 0.37 & 39.17 \\
        \hline
        Total & - & 96 & 2.78 & 10.28 & 0.11 & 86.82 
    \end{tabular}
    \label{tab:non_comp_matches}
\end{table}

\begin{table}[!ht]
    \centering
    \caption{Overview of each type of match per round/slot in the iRR format}
    \begin{tabular}{c|cccccc}
        Round & Slot & \#Matches &  Stakeless (\%) & Asymmetric (\%) & Collusive (\%) & Competitive (\%) \\
        \hline
        7 & 1 & 2 & 0.06 & 4.15 & 0.0 & 95.79 \\
        7 & 2 & 7 & 0.16 & 5.87 & 0.0 & 93.97 \\
        7 & 3 & 2 & 0.6 & 12.22 & 0.0 & 87.18 \\
        7 & 4 & 7 & 0.8 & 13.78 & 0.0 & 85.42 \\
        8 & 1 & 18 & 5.55 & 31.69 & 1.34 & 61.42 \\
        \hline
        Total & - & 144 & 0.75 & 5.14 & 0.17 & 93.94 
    \end{tabular}
    \label{tab:non_comp_matches_irr}
\end{table}

From Tables \ref{tab:non_comp_matches} and \ref{tab:non_comp_matches_irr}, we see that only a small percentage of the matches is observed to be collusive. This might give the impression that collusion is not an issue in both formats. However, if we count the number of collusive games within a season, we observe that there is a $21.11\%$ probability that a season contains a collusive game while this is only $10.48\%$ for the group stage format. Indeed, based on the numbers presented in Tables \ref{tab:non_comp_matches} and \ref{tab:non_comp_matches_irr}, the probability that a season in the group stage format contains at least one collusion opportunity is $1-(1-0.0011)^{96} \times 100\% = 10.18\%$, while for the iRR format, this probability is $1-(1-0.0017)^{144} \times 100\% = 21.73\%$. This indicates that collusion poses a non-negligible threat, and that the iRR format is associated with a noticeable increase in its probability.

In Table \ref{tab:collusion_iRR_decomposed}, we decompose collusion in the iRR format based on the desired prize of the colluding teams. For example, Table \ref{tab:collusion_iRR_decomposed} reveals that in 13.34\% of the observed collusion opportunities, a tie would guarantee the top 8 for both teams. Such a decomposition is not necessary for the group stage format, since the only collusion opportunity that is possible in this format is if two teams guarantee finishing in the top two positions by a tie.

\begin{table}[!ht]
  \centering
  \caption{Observed percentage of the simulated seasons with a collusion opportunity occurring}
  \begin{tabular}{c|c}
    Collusion & Number (\%) \\
    \hline
    T8$^{1}$   & 13.34 \\
    T24$^{2}$  & 32.73 \\
    T8-24$^{3}$& 35.21 \\
    SL$^{4}$   & 18.72 
  \end{tabular}

  \vspace{0.5em}
  \begin{minipage}{\textwidth}
    \footnotesize
    \textsuperscript{1} T8: A tie guarantees top 8 for both teams;  
    \textsuperscript{2} T24: A tie guarantees top 24 for both teams;  
    \textsuperscript{3} T8-24: A tie guarantees top 8 for one and top 24 for the other team;   
    \textsuperscript{4} SL: Small loss.
  \end{minipage}
  
  \label{tab:collusion_iRR_decomposed}
\end{table}

Finally, we show the observed probability of a team in a certain position securing its prize before the last round for the group stage format (Table~\ref{tab:slot_prize_fixed_gs}) and the iRR format (Table~\ref{tab:slot_prize_fixed_iRR}). In both formats, we never observed that a team's prize was fixed before the last two rounds. We observe that the strongest and weakest teams are typically able to secure their prize well before the tournament ends. By contrast, the middle-ranked teams are least likely to know their outcome in advance, as they must both be eliminated from the top 8 and guaranteed a place within the top 24. A similar analysis holds for the third strongest team in each group of the group stage format. Note that the numbers are lower in the odd time slots since only 2 instead of 7 games are played here, hence less information can be revealed in these time slots about which teams can be sure of which prizes. Further note that we cannot simply sum these numbers as the identity of the team in a certain position changes over the time slots.

\begin{table}[!ht]
    \centering
    \caption{Observed probability that a team's prize was secured after a certain time slot in the group stage format}
    \begin{tabular}{c|ccc}
        & \multicolumn{3}{c}{Round} \\
        \hline
        Position & 3 & 4 & 5 \\
        \hline
        1 & 0.0 & 31.92 & 49.46 \\
        2 & 0.0 & 2.57 & 25.17 \\
        3 & 0.0 & 0.0 & 6.44 \\
        4 & 0.0 & 1.97 & 32.29 
\end{tabular}
    \label{tab:slot_prize_fixed_gs}
\end{table}
 
\begin{table}[!ht]
    \centering
    \caption{Observed probability that a team's prize was secured after a certain time slot in the iRR format}
    \begin{tabular}{c|cccccccc}
         & \multicolumn{8}{c}{Round (Time slot)} \\
        \hline
        Position & 6(1) & 6(2) & 6(3) & 6(4) & 7(1) & 7(2) & 7(3) & 7(4) \\
        \hline
        1 & 0.88 & 8.07 & 3.98 & 17.12 & 12.41 & 36.87 & 10.08 & 25.93 \\
        2 & 0.01 & 0.47 & 0.37 & 2.86 & 2.87 & 21.0 & 8.35 & 31.98\\
        3 & 0.0 & 0.01 & 0.01 & 0.29 & 0.46 & 7.07 & 3.8 & 24.99\\
        4 & 0.0 & 0.0 & 0.0 & 0.02 & 0.05 & 1.49 & 1.08 & 13.84\\
        5 & 0.0 & 0.0 & 0.0 & 0.0 & 0.0 & 0.25 & 0.22 & 5.31\\
        6 & 0.0 & 0.0 & 0.0 & 0.0 & 0.0 & 0.02 & 0.03 & 1.37\\
        7 & 0.0 & 0.0 & 0.0 & 0.0 & 0.0 & 0.0 & 0.0 & 0.23\\
        8 & 0.0 & 0.0 & 0.0 & 0.0 & 0.0 & 0.0 & 0.0 & 0.02\\
        9 & 0.0 & 0.0 & 0.0 & 0.0 & 0.0 & 0.0 & 0.0 & 0.01\\
        10 & 0.0 & 0.0 & 0.0 & 0.0 & 0.0 & 0.02 & 0.01 & 0.1\\
        11 & 0.0 & 0.0 & 0.0 & 0.0 & 0.0 & 0.05 & 0.07 & 0.49\\
        12 & 0.0 & 0.0 & 0.0 & 0.0 & 0.0 & 0.11 & 0.17 & 1.4\\
        13 & 0.0 & 0.0 & 0.0 & 0.0 & 0.0 & 0.15 & 0.25 & 3.1\\
        14 & 0.0 & 0.0 & 0.0 & 0.0 & 0.0 & 0.14 & 0.23 & 4.89\\
        15 & 0.0 & 0.0 & 0.0 & 0.0 & 0.0 & 0.1 & 0.22 & 6.19\\
        16 & 0.0 & 0.0 & 0.0 & 0.0 & 0.0 & 0.07 & 0.17 & 6.57\\
        17 & 0.0 & 0.0 & 0.0 & 0.0 & 0.0 & 0.05 & 0.12 & 5.98\\
        18 & 0.0 & 0.0 & 0.0 & 0.0 & 0.0 & 0.02 & 0.06 & 4.62\\
        19 & 0.0 & 0.0 & 0.0 & 0.0 & 0.0 & 0.01 & 0.03 & 2.78\\
        20 & 0.0 & 0.0 & 0.0 & 0.0 & 0.0 & 0.0 & 0.01 & 1.32\\
        21 & 0.0 & 0.0 & 0.0 & 0.0 & 0.0 & 0.0 & 0.0 & 0.44\\
        22 & 0.0 & 0.0 & 0.0 & 0.0 & 0.0 & 0.0 & 0.0 & 0.11\\
        23 & 0.0 & 0.0 & 0.0 & 0.0 & 0.0 & 0.0 & 0.0 & 0.02\\
        24 & 0.0 & 0.0 & 0.0 & 0.0 & 0.0 & 0.0 & 0.0 & 0.0\\
        25 & 0.0 & 0.0 & 0.0 & 0.0 & 0.0 & 0.0 & 0.0 & 0.0\\
        26 & 0.0 & 0.0 & 0.0 & 0.0 & 0.0 & 0.0 & 0.0 & 0.02\\
        27 & 0.0 & 0.0 & 0.0 & 0.0 & 0.0 & 0.01 & 0.01 & 0.15\\
        28 & 0.0 & 0.0 & 0.0 & 0.0 & 0.0 & 0.08 & 0.1 & 0.77\\
        29 & 0.0 & 0.0 & 0.0 & 0.0 & 0.04 & 0.43 & 0.31 & 2.73\\
        30 & 0.0 & 0.0 & 0.0 & 0.0 & 0.18 & 1.81 & 1.1 & 6.94\\
        31 & 0.0 & 0.0 & 0.0 & 0.01 & 0.72 & 5.37 & 2.76 & 14.0\\
        32 & 0.0 & 0.0 & 0.01 & 0.07 & 2.12 & 12.96 & 5.42 & 22.39\\
        33 & 0.0 & 0.02 & 0.02 & 0.43 & 5.2 & 24.42 & 8.26 & 28.06\\
        34 & 0.0 & 0.07 & 0.1 & 1.96 & 10.37 & 36.86 & 9.85 & 27.31\\
        35 & 0.02 & 0.77 & 0.61 & 6.35 & 16.19 & 45.43 & 9.1 & 19.95\\
        36 & 1.02 & 9.12 & 4.16 & 19.31 & 18.14 & 38.97 & 5.53 & 9.85
\end{tabular}
    \label{tab:slot_prize_fixed_iRR}
\end{table}

\subsection{Effect of the schedule in the group stage format}

Next, we analyze the effect of the schedule on the occurrence of the different sates of matches in the group stage format. The observed percentage of stakeless, asymmetric, collusive and competitive matches, averaged over the seasons, is shown in Table~\ref{tab:schedules_group_stage}. The best and worst schedules are indicated in bold.

\begin{table}[!ht]
    \centering
    \caption{Observed percentage of each type of match in the group stage format, averaged over the seasons, for each schedule}
    \begin{tabular}{c|cccc}
         Schedule & Stakeless (\%) & Asymmetric (\%)  & Collusive (\%)  & Competitive (\%) \\
        \hline
        Random & 2.78 & 10.28 & 0.11 & 86.83 \\
        1231 & 2.55 & 10.64 & \textbf{0.14} & 86.67 \\
        2113 & 2.61 & 10.7 & \textbf{0.14} & \textbf{86.56} \\
        1241 & 2.63 & 10.58 & \textbf{0.14} & 86.65 \\
        2114 & 2.52 & 10.35 & 0.12 & 87.0 \\
        1321 & 3.32 & 9.76 & 0.1 & 86.82 \\
        3112 & \textbf{3.4} & 9.79 & 0.11 & 86.7 \\
        1341 & 2.44 & 10.65 & \textbf{0.09} & 86.83 \\
        3114 & 2.4 & 10.22 & 0.1 & \textbf{87.29} \\
        1421 & 3.34 & 9.91 & 0.1 & 86.65 \\
        1431 & 2.4 & \textbf{10.79} & \textbf{0.09} & 86.72 \\
        4112 & 3.38 & \textbf{9.68} & 0.11 & 86.84 \\
        4113 & \textbf{2.38} & 10.27 & 0.1 & 87.26
    \end{tabular}
    \label{tab:schedules_group_stage}
\end{table}

From Table \ref{tab:schedules_group_stage}, we see that there is not one schedule that performs best or worst with respect to all types of non-competitive matches. However, if we look at competitive matches, schedule 3114 performs best, while schedule 2113 performs worst. The Friedman's test reveals that there are significant differences between the schedules, for each type of match (p < 0.01). Moreover, according to the Nemenyi test, the best and worst schedules for each type of match significantly differ. However, regarding competitive matches, assigning each of the schedules 1321 (p = 1.0), 1341 (p > 0.99) or 4112 (p > 0.98) to all the groups does not yield significantly different results compared to randomly assigning a schedule to each group. 
The results per season can be found in the Appendix. 

\subsection{Effect of the schedule in the iRR format}

\begin{table}[!ht]
    \centering
    \caption{Observed percentage of each type of match in the iRR format, averaged over the seasons, for each schedule}
    \begin{tabular}{c|cccc}
         Schedule & Stakeless (\%) & Asymmetric (\%) & Collusive (\%) & Competitive (\%) \\
        \hline
        $\mathcal{M}_0$ & 0.75 & 5.14 & 0.17 & 93.94 \\
        $\mathcal{M}_1$ & 0.66 & 4.63 & 0.15 & 94.56 \\
        $\mathcal{M}_2$ & \textbf{1.13} & \textbf{5.28} & \textbf{0.23} & \textbf{93.36} \\
        $\mathcal{M}_3$ & 0.67 & 5.05 & 0.16 & 94.12 \\
        $\mathcal{M}_4$ & \textbf{0.58} & \textbf{4.1} & \textbf{0.13} & \textbf{95.19} 
    \end{tabular}
    \label{tab:schedules_irr}
\end{table}

Finally, we turn to the effect of the schedule in the iRR format. The Friedman's test reveals that the schedules significantly differ for each type of match (p < 0.01). Moreover, the Nemenyi test shows that, for each type of match, every pair of schedules differs significantly (p < 0.01). Table~\ref{tab:schedules_irr} shows that schedule $\mathcal{M}_2$, in which matches between teams of similar strength are scheduled in the last matchdays performs worst. A possible explanation for this result is that with this scheduling strategy, the stronger teams play all their games against weaker teams in the beginning of the competition. As a result, the stronger teams are more likely to already have obtained enough points to secure their prize before the last matchdays. Table \ref{tab:schedules_irr} further suggests that the opposite strategy ($\mathcal{M}_3$) does slightly perform better than a random schedule. However, schedule $\mathcal{M}_4$, which a) avoids that teams start or end with two matches against two teams from pot 1 or 2 and b) avoids that teams start or end with two matches against two teams from pot 3 or 4, performs best. This suggests that balancing matches of the stronger teams over the competition, rather than concentrating them at the beginning or end, has a favorable effect on competitiveness. Notably, the schedule which maximizes television audience yields the second best results. This is encouraging, as it suggests that attractiveness from a commercial as well as from a competitive viewpoint are not conflicting with each other. The results per season can be found in the Appendix.

\section{Conclusion}\label{sec:conclusion}

This paper compares the occurrence of different types of non-competitive matches under the former group stage and the newly introduced incomplete round-robin (iRR) format of the UEFA Champions League. non-competitive matches are identified through a mathematically exact classification, and a simulation study is employed to evaluate the two formats. The findings indicate that, across a range of plausible scheduling scenarios, the iRR format substantially reduces the proportion of stakeless and asymmetric matches. Furthermore, appropriate scheduling strategies can further decrease the prevalence of non-competitive games. In particular, balancing fixtures against strong and weak opponents enhances competitiveness. Notably, maximizing competitiveness and optimizing television audiences appear to be mutually compatible objectives. Conversely, scheduling late-stage matches between teams of similar strength increases the likelihood of non-competitive outcomes and is therefore undesirable.

Previous research on collusive matches has shown that opportunities for collusion are considerably more prevalent when the final games are not played simultaneously. In such cases, teams playing later gain perfect information about the minimum result required for qualification, making strategic outcomes such as small-loss collusion more likely. Conversely, scheduling all final-round matches at the same time reduces these opportunities but imposes costs in terms of television viewership. A promising direction for future research is to quantify this trade-off by examining how different scheduling configurations for the last round affect both the incidence of match manipulation and the associated commercial implications.

We conclude that UEFA’s statement that the new format ``results in more competitive matches'' is grounded in facts. On a more detailed level, when distinguishing  non-competitive matches in stakeless, asymmetric, and collusive, we find that in the new format, the number of stakeless and asymmetric matches decreases significantly, while there is a small increase in collusive matches.


\section*{Acknowledgments}
  The computational resources (Stevin Supercomputer Infrastructure) and services used in this work were provided by the VSC (Flemish Supercomputer Center), funded by Ghent University, FWO and the Flemish Government – department EWI. We also wish to thank Andr{\'a}s Gyimesi and Roel Lambers for fruitful discussions on the Champions League reform and L{\'a}szl{\'o} Csat{\'o} in particular for his helpful suggestions on the writing of this paper.

\clearpage 

\section*{Appendix}

\subsection*{Results decomposed per season}

\begin{table}[!ht]
    \centering
    \caption{Percentage of games that are stakeless, for each schedule for each season in the group stage format.}
    \begin{tabular}{c|ccccccc}
        Schedule & 2019-2020 & 2020-2021 & 2021-2022 & 2022-2023 & 2023-2024 & 2024-2025 & Average \\
        \hline
        Random & 2.6 & 2.88 & 3.07 & 2.72 & 2.61 & 2.81 & 2.78 \\
        1231 & 2.36 & 2.72 & 2.68 & 2.59 & 2.37 & 2.61 & 2.55 \\
        2113 & 2.42 & 2.81 & 2.8 & 2.7 & 2.34 & 2.59 & 2.61 \\
        1241 & 2.48 & 2.72 & 3.03 & 2.56 & 2.35 & 2.62 & 2.63 \\
        2114 & 2.36 & 2.59 & 2.88 & 2.43 & 2.27 & 2.6 & 2.52 \\
        1321 & 3.07 & 3.34 & 3.68 & 3.16 & 3.36 & 3.29 & 3.32 \\
        3112 & \textbf{3.21} & \textbf{3.47} & 3.73 & \textbf{3.25} & \textbf{3.42} & 3.35 & \textbf{3.4} \\
        1341 & 2.3 & 2.56 & 2.76 & 2.38 & 2.14 & 2.48 & 2.44 \\
        3114 & 2.27 & \textbf{2.49} & 2.73 & \textbf{2.34} & \textbf{2.1} & 2.46 & 2.4 \\
        1421 & 3.09 & 3.35 & 3.7 & 3.18 & 3.39 & 3.32 & 3.34 \\
        1431 & \textbf{2.22} & 2.58 & 2.56 & 2.48 & 2.13 & 2.44 & 2.4 \\
        4112 & 3.19 & 3.39 & \textbf{3.74} & 3.18 & 3.4 & \textbf{3.37} & 3.38 \\
        4113 & 2.24 & 2.54 & \textbf{2.5} & 2.47 & \textbf{2.1} & \textbf{2.42} & \textbf{2.38} 
    \end{tabular}
    \label{tab:gs_S}
\end{table} 

\begin{table}[!ht]
    \centering
    \caption{Percentage of games that are asymmetric, for each schedule for each season in the group stage format.}
    \begin{tabular}{c|ccccccc}
        Schedule & 2019-2020 & 2020-2021 & 2021-2022 & 2022-2023 & 2023-2024 & 2024-2025 & Average \\
        \hline
        Random & 10.3 & 10.34 & 10.28 & 10.27 & 10.18 & 10.31 & 10.28 \\
        1231 & 10.66 & 10.6 & 10.78 & 10.53 & 10.61 & 10.65 & 10.64 \\
        2113 & 10.72 & 10.79 & 10.8 & 10.56 & 10.73 & 10.56 & 10.7 \\
        1241 & 10.44 & 10.66 & 10.42 & 10.61 & 10.63 & \textbf{10.75} & 10.58 \\
        2114 & 10.3 & 10.44 & 10.2 & 10.36 & 10.36 & 10.46 & 10.35 \\
        1321 & 9.85 & 9.85 & \textbf{9.62} & 9.83 & 9.46 & 9.95 & 9.76 \\
        3112 & 9.8 & 9.94 & 9.87 & 9.98 & 9.45 & \textbf{9.7} & 9.79 \\
        1341 & 10.55 & 10.71 & 10.55 & \textbf{10.65} & 10.72 & 10.72 & 10.65 \\
        3114 & 10.15 & 10.31 & 10.05 & 10.23 & 10.25 & 10.3 & 10.22 \\
        1421 & 10.03 & 10.1 & 9.95 & 10.08 & \textbf{9.42} & 9.89 & 9.91 \\
        1431 & \textbf{10.86} & \textbf{10.85} & \textbf{10.92} & \textbf{10.65} & \textbf{10.79} & 10.69 & \textbf{10.79} \\
        4112 & \textbf{9.64} & \textbf{9.72} & 9.68 & \textbf{9.73} & 9.48 & 9.8 & \textbf{9.68} \\
        4113 & 10.26 & 10.31 & 10.36 & 10.16 & 10.21 & 10.29 & 10.27 
    \end{tabular}
    \label{tab:gs_A}
\end{table} 

\begin{table}[!ht]
    \centering
    \caption{Percentage of games that are collusive, for each schedule for each season in the group stage format.}
    \begin{tabular}{c|ccccccc}
        Schedule & 2019-2020 & 2020-2021 & 2021-2022 & 2022-2023 & 2023-2024 & 2024-2025 & Average \\
        \hline
        Random & 0.11 & 0.11 & 0.11 & 0.12 & 0.11 & 0.11 & 0.11 \\
        1231 & 0.13 & 0.13 & \textbf{0.14} & \textbf{0.14} & \textbf{0.15} & \textbf{0.12} & \textbf{0.14} \\
        2113 & \textbf{0.14} & \textbf{0.14} & \textbf{0.14} & \textbf{0.14} & 0.14 & \textbf{0.12} & \textbf{0.14} \\
        1241 & 0.13 & \textbf{0.14} & \textbf{0.14} & \textbf{0.14} & \textbf{0.15} & \textbf{0.12} & \textbf{0.14} \\
        2114 & 0.12 & 0.12 & 0.12 & 0.12 & 0.14 & \textbf{0.12} & 0.12 \\
        1321 & 0.1 & 0.11 & 0.11 & 0.11 & 0.11 & 0.09 & 0.1 \\
        3112 & 0.12 & 0.11 & 0.1 & 0.11 & 0.1 & 0.1 & 0.11 \\
        1341 & 0.09 & 0.1 & \textbf{0.09} & \textbf{0.09} & \textbf{0.08} & \textbf{0.08} & \textbf{0.09} \\
        3114 & 0.1 & 0.1 & \textbf{0.09} & 0.11 & 0.09 & 0.09 & 0.1 \\
        1421 & 0.1 & 0.1 & 0.11 & 0.1 & 0.1 & 0.09 & 0.1 \\
        1431 & \textbf{0.08} & \textbf{0.09} & \textbf{0.09} & \textbf{0.09} & 0.09 & \textbf{0.08} & \textbf{0.09} \\
        4112 & 0.11 & 0.1 & 0.11 & 0.11 & 0.11 & 0.1 & 0.11 \\
        4113 & 0.1 & 0.1 & 0.1 & 0.1 & 0.1 & 0.09 & 0.1
    \end{tabular}
    \label{tab:gs_C}
\end{table} 

\begin{table}[!ht]
    \centering
    \caption{Percentage of games that are competitive, for each schedule for each season in the group stage format.}
    \begin{tabular}{c|ccccccc}
        Schedule & 2019-2020 & 2020-2021 & 2021-2022 & 2022-2023 & 2023-2024 & 2024-2025 & Average \\
        \hline
        Random & 86.99 & 86.67 & 86.54 & 86.89 & 87.09 & 86.78 & 86.83 \\
        1231 & 86.84 & 86.55 & 86.41 & 86.75 & 86.88 & 86.62 & 86.67 \\
        2113 & \textbf{86.73} & \textbf{86.25} & 86.25 & \textbf{86.6} & \textbf{86.78} & 86.72 & \textbf{86.56} \\
        1241 & 86.94 & 86.49 & 86.41 & 86.69 & 86.87 & \textbf{86.52} & 86.65 \\
        2114 & 87.21 & 86.84 & 86.8 & 87.09 & 87.23 & 86.82 & 87.0 \\
        1321 & 86.97 & 86.71 & 86.6 & 86.9 & 87.07 & 86.67 & 86.82 \\
        3112 & 86.88 & 86.48 & 86.3 & 86.65 & 87.03 & 86.85 & 86.7 \\
        1341 & 87.06 & 86.63 & 86.6 & 86.88 & 87.06 & 86.72 & 86.83 \\
        3114 & \textbf{87.48} & \textbf{87.1} & \textbf{87.13} & \textbf{87.32} & 87.56 & 87.15 & \textbf{87.29} \\
        1421 & 86.78 & 86.45 & \textbf{86.24} & 86.65 & 87.09 & 86.7 & 86.65 \\
        1431 & 86.84 & 86.48 & 86.43 & 86.78 & 86.99 & 86.78 & 86.72 \\
        4112 & 87.05 & 86.79 & 86.47 & 86.99 & 87.02 & 86.72 & 86.84 \\
        4113 & 87.41 & 87.05 & 87.04 & 87.27 & \textbf{87.59} & \textbf{87.2} & 87.26 
    \end{tabular}
    \label{tab:gs_uncomp}
\end{table} 

\begin{table}[!ht]
    \centering
    \caption{Percentage of games that are stakeless, for each schedule for each season in the iRR format.}
    \begin{tabular}{c|ccccccc}
        Schedule & 2019-2020 & 2020-2021 & 2021-2022 & 2022-2023 & 2023-2024 & 2024-2025 & Average \\
        \hline
        $\mathcal{M}_0$ & 0.72 & 0.78 & 0.82 & 0.7 & 0.7 & 0.77 & 0.75 \\
        $\mathcal{M}_1$ & \textbf{0.41} & 0.71 & 0.8 & 0.65 & 0.67 & 0.71 & 0.66 \\
        $\mathcal{M}_2$ & \textbf{1.08} & \textbf{1.19} & \textbf{1.3} & \textbf{1.0} & \textbf{0.98} & \textbf{1.22} & \textbf{1.13} \\
        $\mathcal{M}_3$ & 0.64 & 0.69 & 0.74 & 0.63 & 0.65 & 0.68 & 0.67 \\
        $\mathcal{M}_4$ & 0.55 & \textbf{0.58} & \textbf{0.66} & \textbf{0.52} & \textbf{0.55} & \textbf{0.61} & \textbf{0.58}
    \end{tabular}
    \label{tab:irr_S}
\end{table} 

\begin{table}[!ht]
    \centering
    \caption{Percentage of games that are asymmetric, for each schedule for each season in the iRR format.}
    \begin{tabular}{c|ccccccc}
        Schedule & 2019-2020 & 2020-2021 & 2021-2022 & 2022-2023 & 2023-2024 & 2024-2025 & Average \\
        \hline
        $\mathcal{M}_0$ & 5.04 & 5.12 & 5.5 & 4.93 & 4.96 & 5.31 & 5.14 \\
        $\mathcal{M}_1$ & \textbf{3.05} & 4.91 & 5.31 & 4.73 & 4.83 & 4.98 & 4.63 \\
        $\mathcal{M}_2$ & \textbf{5.17} & \textbf{5.32} & \textbf{5.54} & \textbf{5.13} & \textbf{5.18} & \textbf{5.34} & \textbf{5.28} \\
        $\mathcal{M}_3$ & 4.99 & 5.03 & 5.38 & 4.87 & 4.85 & 5.18 & 5.05 \\
        $\mathcal{M}_4$ & 4.05 & \textbf{4.1} & \textbf{4.38} & \textbf{3.93} & \textbf{3.96} & \textbf{4.19} & \textbf{4.1}
    \end{tabular}
    \label{tab:irr_A}
\end{table} 

\begin{table}[!ht]
    \centering
    \caption{Percentage of games that are collusive, for each schedule for each season in the iRR format.}
    \begin{tabular}{c|ccccccc}
        Schedule & 2019-2020 & 2020-2021 & 2021-2022 & 2022-2023 & 2023-2024 & 2024-2025 & Average \\
        \hline
        $\mathcal{M}_0$ & 0.16 & 0.18 & 0.17 & 0.16 & 0.16 & 0.17 & 0.17 \\
        $\mathcal{M}_1$ & \textbf{0.09} & 0.16 & 0.17 & 0.15 & 0.15 & 0.15 & 0.15 \\
        $\mathcal{M}_2$ & \textbf{0.22} & \textbf{0.23} & \textbf{0.24} & \textbf{0.22} & \textbf{0.21} & \textbf{0.23} & \textbf{0.23} \\
        $\mathcal{M}_3$ & 0.15 & 0.16 & 0.16 & 0.15 & 0.15 & 0.16 & 0.16 \\
        $\mathcal{M}_4$ & 0.12 & \textbf{0.14} & \textbf{0.14} & \textbf{0.12} & \textbf{0.13} & \textbf{0.13} & \textbf{0.13}
    \end{tabular}
    \label{tab:irr_C}
\end{table} 

\begin{table}[!ht]
    \centering
    \caption{Percentage of games that are competitive, for each schedule for each season in the iRR format.}
    \begin{tabular}{c|ccccccc}
        Schedule & 2019-2020 & 2020-2021 & 2021-2022 & 2022-2023 & 2023-2024 & 2024-2025 & Average \\
        \hline
        $\mathcal{M}_0$ & 94.09 & 93.92 & 93.5 & 94.21 & 94.17 & 93.74 & 93.94 \\
        $\mathcal{M}_1$ & \textbf{96.45} & 94.22 & 93.72 & 94.46 & 94.35 & 94.16 & 94.56 \\
        $\mathcal{M}_2$ & \textbf{93.53} & \textbf{93.26} & \textbf{92.92} & \textbf{93.65} & \textbf{93.63} & \textbf{93.2} & \textbf{93.36} \\
        $\mathcal{M}_3$ & 94.22 & 94.12 & 93.72 & 94.34 & 94.35 & 93.98 & 94.12 \\
        $\mathcal{M}_4$ & 95.28 & \textbf{95.18} & \textbf{94.81} & \textbf{95.43} & \textbf{95.35} & \textbf{95.06} & \textbf{95.19} \\
    \end{tabular}
    \label{tab:irr_uncomp}
\end{table} 

\clearpage

\begingroup
\setstretch{1.1}
\setlength{\bibsep}{0pt plus 0.3ex}
\small{\bibliography{bibliography}}
\endgroup

\end{document}